\documentclass[12pt]{revtex4}
\usepackage{graphicx}
\usepackage[english]{babel}
\begin{document}
\title{Manipulation of the graphene surface potential by ion irradiation}

\author{O.~Ochedowski$^1$, B.~Kleine Bussmann$^1$, B.~Ban~d'Etat$^2$, H.~Lebius$^2$ and M.~Schleberger$^1$ \footnote{electronic address: marika.schleberger@uni-due.de}} 
\affiliation{$^1$Fakult\"at f\"ur Physik and CeNIDE, Universit\"at Duisburg-Essen, D-47048 Duisburg, Germany}
\affiliation{$^2$CIMAP (CEA-CNRS-ENSICAEN-UCBN), 14070 Caen Cedex 5, France}

\begin{abstract}
We show that the work function of exfoliated single layer graphene can be modified by irradiation with swift ($E_{kin}=92$~MeV) heavy ions under glancing angles of incidence. Upon ion impact individual surface tracks are created in graphene on SiC. Due to the very localized energy deposition characteristic for ions in this energy range, the surface area which is structurally altered is limited to $\approx 0.01~\mu$m$^2$ per track. Kelvin probe force microscopy reveals that those surface tracks consist of electronically modified material and that a few tracks suffice to shift the surface potential of the whole single layer flake by $\approx 400$~meV. Thus, the irradiation turns the initially $n$-doped graphene into $p$-doped graphene with a hole density of $8.5\times10^{12}$~holes/cm$^2$. This doping effect persists even after heating the irradiated samples to 500$^{\mathrm{\circ}}$C. Therefore, this charge transfer is not due to adsorbates but must instead be attributed to implanted atoms. The method presented here opens up a new way to efficiently manipulate the charge carrier concentration of graphene.

\end{abstract}

\maketitle
Modern electronics began with the realisation of the first bipolar and field effect transistors in the late 1940s  \cite{SHOCKLEY.1948,Dormael.2009}. A technologically important step was the development of efficient doping techniques in the late 1970s as they are a prerequisite to realise e.g.~microscopic $p$--$n$ junctions. Bulk semiconductors like silicon are often doped by ion implanation. This technique relies on well defined projected ranges due to binary collisions and allows to control the electron and hole concentration in e.g.~silicon in a very large range from $10^{13}$/cm$^{3}$ up to $10^{21}$/cm$^{3}$ \cite{Doering.2008,Ziegler.1984}. Nowadays graphene has attracted a lot of interest as an excellent candidate for future nanoelectronics due to its promising electronic properties \cite{Geim.2007,Geim.2009}. Especially single layer graphene (SLG) is known to exhibit a very high charge carrier mobility \cite{Zhang.2005,Park.2008}. For electronic applications though it is crucial to find a way to alter the type and concentration of the charge carriers. Because of its 2D nature classical ion implantation is not feasible for SLG. Research on alternative doping mechanisms for graphene is therefore of utmost importance.

There are several approaches to achieve controlled doping of graphene which have already been investigated: One is chemical doping by using adsorbates which is realised by deposition of molecules or coating on top of graphene sheets \cite{Brenner.2010}. Another approach is using the substrate to induce charge transfer at the interface. This has already been investigated for insulating \cite{Shi.2009,Bussmann.2011,Bukowska.2011} as well as for metallic substrates \cite{Giovannetti.2008}. Doping by keV ion implantation has been studied theoretically \cite{Zhao.2012b} and it was shown that the substrate enhances the chances for successfull indirect implantation. Experimentally, ion implantation in epitaxial graphene grown on SiC as well as on Ni(111) has been performed by irradiation with slow (keV) nitrogen ions at high fluences \cite{Kim.2010,Zhao.2012}. In freestanding graphene single atom substitution has been demonstrated \cite{Wang.2011}. However, in both reports the samples were not investigated with respect to changes of the carrier concentration. In this report we show that a manipulation of the graphene surface potential is feasible by irradiations with swift heavy ions (SHI) at low fluences. 

For this experiment graphene was exfoliated from bulk graphite (HOPG - Advanced Ceramics) onto a 6H-SiC(0001) substrate (Pam-Xiamen) in ambient conditions without prior treatment of the substrate. In order to unambiguously identify single layer graphene (SLG) Raman spectroscopy measurements were performed \cite{Ferrari.2006}. 

\begin{figure}[ht!]
    \centering
\includegraphics[width=0.8\textwidth]{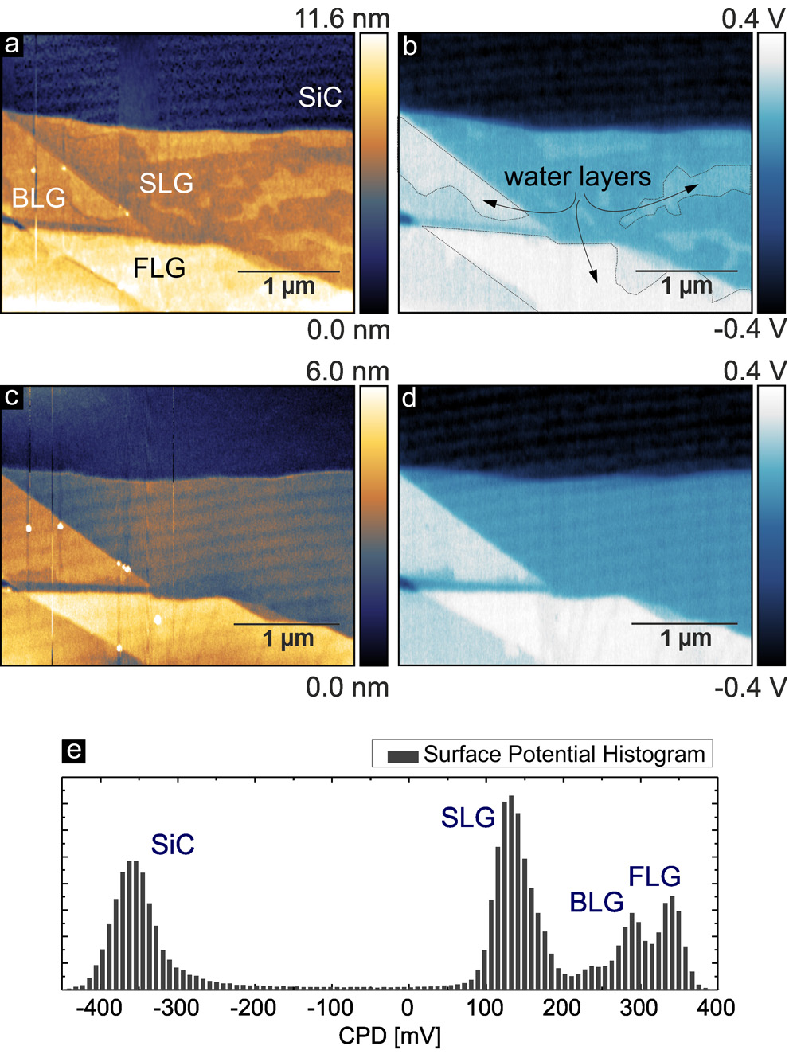}
    \caption{AFM ((a) measured at $df=-4$~Hz, (c) measured at $df=-10$~Hz) and KPFM (b),(d) images of samples as prepared (a),(b) and after thermal processing (c),(d) measured in UHV. From the histogram (e) it can bee seen that the surface potential of SLG and BLG after thermal processing is shifted downwards and decreases with decreasing layer thickness, indicating $n$-type doping (see text).}
    \label{figure1}
\end{figure}

To modify the electronic structure of graphene by SHI irradiation it is crucial to remove any intercalated water between graphene and the substrate material. Failure to remove this water layer will result in the creation of SHI induced foldings which have been described in detail elsewhere \cite{Akcoltekin.2011b}. Therefore, the samples were introduced into an ultra high vacuum system (UHV, base pressure $p_b=10^{-10}$~mbar) for cleaning by thermal processing (see below). The removal of water layers and other adsorbates at the graphene substrate interface was observed by non-contact atomic force microscopy (AFM ) and simultaneous Kelvin probe force microscopy (KPFM) in UHV, see fig.~\ref{figure1}. The KPFM measurements to obtain surface potential maps were performed using a RHK UHV 7500 system and the PLL Pro 2 controller with the inbuilt lock-in by applying an AC voltage of $U_{AC}\approx1$~V and $f\approx1$~kHz to the AFM tip \cite{Nonnenmacher.1991}. In addition, a DC voltage was applied to the AFM tip to minimize the resulting electrostatic forces. The sample was grounded for these measurements. Standard Si-cantilever (Vistaprobe T300) with typical resonance frequencies of about 300~kHz were used. 

In figs.\ref{figure1}(a) and (b) pristine graphene flake with different numbers of layers directly after transfer into the UHV is shown. The topography image shown in fig.\ref{figure1}(a) reveals that water layers with a typical height of 0.5~nm are present at the interface between SLG and the SiC substrate. In the corresponding surface potential map (fig.\ref{figure1}(b)) it can be observed that these water layers increase the contact potential difference (CPD) by roughly 40~mV with respect to the clean graphene substrate system. A similar effect can be seen in bilayer graphene (BLG) and few layer graphene (FLG) sheets. Here, however, the exact position of the water layers is unkown. The water adlayers can be located at the interface between graphene sheets and the SiC substrate or intercalated between graphene sheets.

By repeatedly heating this sample to over 500$^{\mathrm{\circ}}$C for more than one hour, the interfacial water layer between SLG and the substrate can be removed as shown in figs.~\ref{figure1}(c) and (d). The surface potential is now uniformly distributed over the complete SLG sheet. In contrast, water layers in BLG and FLG cannot be completely removed by this procedure. In FLG the water layer has been diminished but BLG is almost completely intercalated (a small area devoid of water can still be seen at the lower edge of the flake). In fig.~\ref{figure1}(e) a histogram of the surface potential map is shown which reveals a clear dependency of the surface potential on the number of graphene layers. Before irradiation, the surface potential clearly {\it increases} with increasing layer number.

After the interfacial water layer between SLG and the SiC substrate had been removed, the samples were transferred in ambient conditions to the IRRSUD beamline for irradiation. The samples were irradiated with 92 MeV $^{129}$Xe$^{23+}$ under a glancing angle of incidence of $\Theta=0.3^{\mathrm{\circ}}$ with respect to the sample surface. The ion fluence was adjusted to result in approximately 3 to 4 surface tracks per $\mu$m$^2$, assuming one track per incident ion \cite{Karlusic.2010}. 

\begin{figure}[ht!]
    \centering
\includegraphics[width=\textwidth]{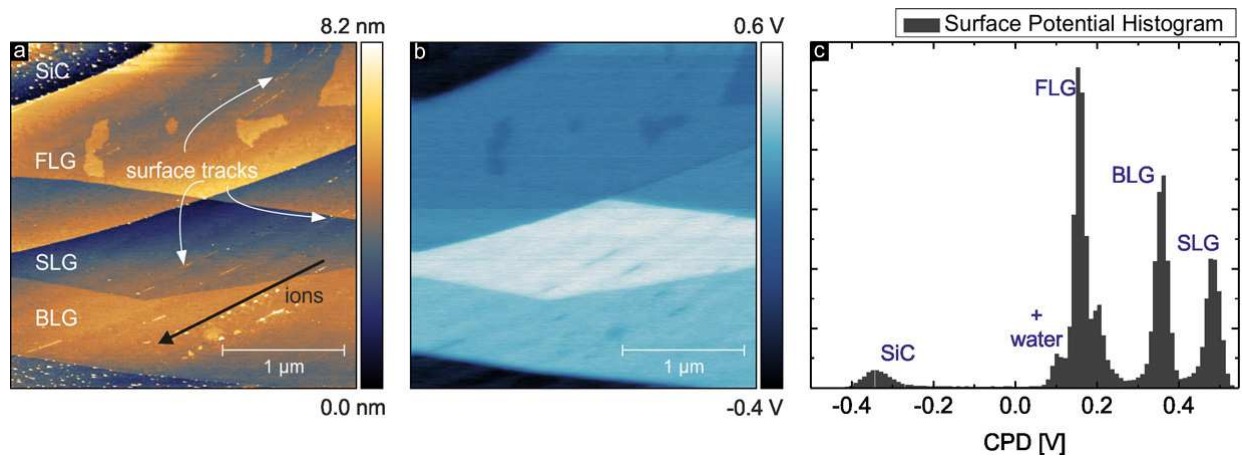}
   	  \caption{AFM ((a) measured at $df=-20$~Hz) and KPFM (b) images of samples after irradiation with Xe ions with a kinetic energy of 93~MeV. Direction of the incoming ions marked by the black arrow. Surface tracks can be observed and the surface potential of SLG and BLG is shifted upwards. It decreases with decreasing layer thickness, see histogram (c), which indicates $p$-type doping. In comparison to the histogram in fig.~\ref{figure1}(c), the layer dependency of the surface potential is inverted.}
    \label{figure2}
\end{figure}

The effect of the SHI irradiation can be seen in the AFM topography image and the surface potential map shown in fig.~\ref{figure2}.
The AFM topography (fig.~\ref{figure2}a) shows surface tracks in form of elongated protrusions on all graphene sheets. The average length of the tracks is $(860\pm290)$~nm. Apart from these modifications SLG and BLG show no signs of defects due to irradiation. The effect of SHI irradiation on the surface potential of graphene is by far more profound as can be seen in figs.~\ref{figure2}(b) and (c). The surface potential shows a contrast inversion with respect to the layer dependency, i.e.~the surface potential now {\it decreases} with increasing layer number. With respect to FLG, the surface potentials of SLG and BLG show an inversion while the surface potential of the SiC substrate is similar to its pristine state.

\begin{figure}[ht!]
    \centering
\includegraphics[width=0.6\textwidth]{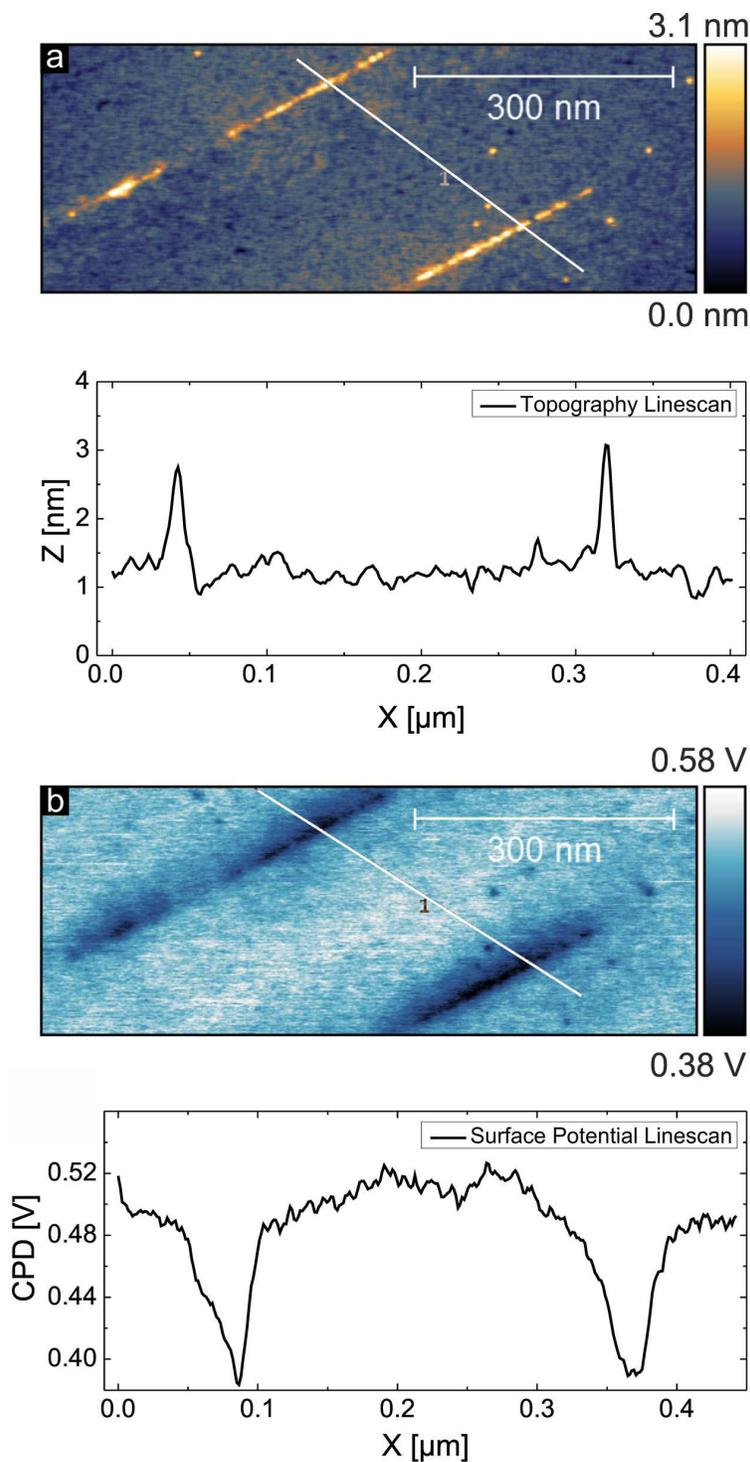}
    \caption{Detailed view of ion induced surface tracks in SLG: (a) AFM topography ($df=-24$~Hz) and (b) surface potential with respective line profiles. The SHI irradiation under glancing incidence angle causes the formation of surface tracks in the SLG sheet with an average length of 860~nm and typical height of about 1.25~nm. These surface tracks locally decrease the surface potential significantly by $\approx 100$~meV.}
    \label{figure3}
\end{figure}

In fig.~\ref{figure3} a detailed view of a surface track on SLG and the corresponding surface potential are shown. These tracks consist of a chain of randomly distributed protrusions with an average height of about 1.25~nm, a width of 10~nm and a length of 280 nm and 400~nm, respectively. The surface potential shows a strong variation at the position of the tracks. 

\begin{figure}[ht!]
    \centering
\includegraphics[width=0.9\textwidth]{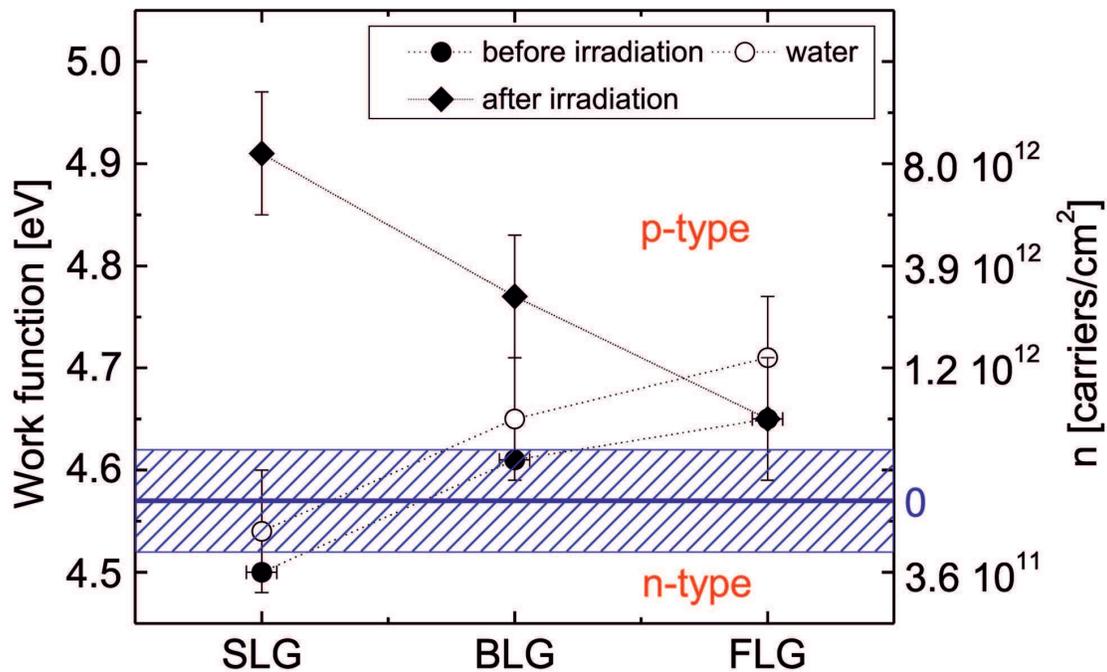}
    \caption{Work function (lefthand scale) and charge carrier density (righthand scale) for graphene of varying layer number determined from samples before irradiation and thermal processing (open circles), before irradiation and after thermal processing (full circles) and after irradiation (diamonds). Lines to guide the eye. The hatched region marks the values reported for undoped graphene.}
    \label{figure4}
\end{figure}

To visualize these results more clearly, we calibrate our KPFM measurements by assigning the known work function of HOPG \cite{Bussmann.2011,Shim.2012} to FLG. Highly oriented pyrolytic graphite is an inert material with a work function of about $\Phi=4.65$~eV \cite{Ooi.2006}. We thus use the relation $\phi=4.65$~eV$+$~e$(CPD_{FLG}-CPD_{nG})$ to assign work functions to SLG and BLG before and after irradiation. The data is shown in fig.~\ref{figure4}. The work function of pristine SLG on SiC is determined to be 4.50$\pm$0.06~eV. BLG on SiC yields 4.61$\pm$0.06~eV, where intercalated and interfacial water layers further increase the work function by 0.04~eV. As discussed above, a strong shift in the work function can be observed for the irradiated graphene sheets, especially SLG. Irradiated SLG on SiC exhibits a work function of 4.91$\pm$0.06~eV and the work function of irradiated BLG on SiC is 4.77$\pm$0.06~eV.

A work function change can be directly related to the charge carrier concentration in the graphene sheet because it corresponds to a Fermi level shift with respect to the Dirac point in the following way: $n=\frac{1}{\pi}(\frac{\Delta E_{F}}{h\nu_{F}})^{2}$. We assume $\nu_{F}$=1$\times$10$^{6}$$\frac{m}{s}$ for the Fermi velocity and use the value for undoped single layer graphene reported by Yu et al.~of $4.57\pm0.05$~eV. The corresponding values are shown in fig.\ref{figure4}, see scale on the righthand side. From this figure we can directly deduce that prior to irradiation, the SLG is $n$-doped ($3.6\times10^{11}$~electrons/cm$^2$) while after irradiation it is $p$-doped ($8.5\times10^{12}$~holes/cm$^2$). Thus, the net transfer of carriers into SLG amounts to $9\times10^{12}$~holes/cm$^2$. 

Let us first comment on the results for SLG/SiC before irradiation. For exfoliated graphene typically $p$-type doping is reported \cite{Ryu.2010} while $n$-type doping represents an exception \cite{Bukowska.2011}. However, very often $p$-type doping will in fact be due to charge transfer from adsorbate layers like water, oxygen and nitrogen which are always present in samples exfoliated in ambient conditions \cite{Ryu.2010,Levesque.2011}. This can be seen also from the results presented here, where the freshly exfoliated SLG flake shows a reduced $n$-type doping in areas where water layers are present, see fig.~\ref{figure1}(b).

Recent experiments on epitaxially grown graphene on SiC also reported $n$-type doping \cite{Varchon.2007,Filleter.2008} and a similar $\Delta$CPD between SLG and BLG of 130~mV. However, Filleter et al.~proposed the C-rich interface layer between graphene and SiC to be the main reason for the $n$-type doping in SLG \cite{Filleter.2008} which is not present in our exfoliated sample. In our case graphene is most likely $n$-doped by a charge transfer from the underlying SiC substrate (which is covered by a $\approx 1$~nm thick oxide) and water layers present at the interface increase the work function of SLG by 40~mV. The fact that water adlayers can effectively block charge transfer between the substrate and graphene has already been reported \cite{Shim.2012}. Kelvin probe force microscopy and transport measurements performed in ambient on epitaxially grown graphene on SiC showed a much lower doping compared to in situ measurements \cite{Curtin.2011}. This effect was attributed to the exposure to air or passivation in agreement with our results.  

Next, we discuss the global doping of SLG observed after irradiation. First we need to exclude the possibility of work function changes due to mechanisms not related to the irradiation. Because of the transfer and the exposure to the vacuum ($p_b=10^{-6}$~mbar) during the irradiation the samples could have been contaminated, e.g.~by water and carbohydrates. Therefore, after the $n$ to $p$ transition had been observed, the irradiated samples were again thermally processed in our UHV chamber. The KPFM image did not change after this treatment, the SLG remained hole doped. For comparison we used another sample which underwent all the processing steps except for the irradiation itself. On this sample, a transition from $n$-type back to $p$-type doping was observed after exposure to air, but in contrast to the irradiated samples this transition could be reversed by thermal processing. From this we can safely exclude contaminants as the origin for the measured doping after irradiation.

The observed doping of SLG with holes must be related to the irradiation itself. Direct implantation of projectile ions can be excluded, as their projected depth is $\approx 50$~nm \cite{Ziegler.2010} and the fluence would be much too low. The surface tracks which are created by SHI in the SiC are local and far apart, they cannot be responsible for a global change of the insulating substrate. Substrate doping is therefore also not the responsible mechanism. However, because graphene is conducting it could very well be affected globally if the tracks present a sufficient amount of acceptors. Whether this is feasible or not can be estimated as follows: With three tracks per $\mu$m$^2$ and each track being 860~nm long and 10~nm wide, the modified area per $\mu$m$^2$ amounts to $0.26 \times 10^{5}$~nm$^2$ in total. On the other hand, the observed shift in surface potential corresponds to a hole density of $1 \times 10^{5}$~holes/$\mu$m$^2$. This means that five to six graphene unit cells would have to contribute one hole each to achieve the observed level of doping. It is therefore entirely possible that the track regions act as acceptors for electrons thereby depleting the whole SLG of holes.

Now the question arises how the tracks can act as acceptors. We propose that this is due to the incorporation of foreign material into the graphene lattice. The atoms would stem from the underlying substrate. It has already been proposed in a theoretical paper by Zhao et al.~that indirect implantation of graphene might be possible with keV ions \cite{Zhao.2012}. In the case of SHI the ejection of surface material is caused by a different process, e.g.~Coulomb explosion \cite{Fleischer.1965} or a thermal spike \cite{Toulemonde.1992}, but it definitely occurs, especially under glancing angle conditions. In substrates covered by graphene this leads to folding of graphene \cite{Akcoltekin.2011b} but under certain circumstances this can be avoided and then graphene is indeed able to catch sputtered material as has been reported recently \cite{Ochedowski.2012}. The proposed implantation would be fully consistent with the fact that the tracks cannot be removed by heating. 

When the SHI induces defects in the graphene lattice along its trajectory, atoms from the substrate material could replace carbon vacancies. A similar effect has been reported by Wang et al., who deposited different dopants in SLG vancies \cite{Wang.2011}. From the current data it remains unclear which foreign atoms are actually incorporated in the graphene lattice. These would most likely be the constituents of the substrate, i.e.~basically Si (as C is already present in graphene) and O from the oxide layer covering the SiC susbtrate. Because of the limited spatial resolution of the KPFM measurement, a detailed analysis of the chemical composition of the surface track can currently not be performed. It would be interesting to see whether by a variation of the substrate material also SHI induced $n$-type doping can be achieved.

In conclusion, we have shown that the work function of exfoliated graphene on SiC significantly changes upon irradiation with swift heavy ions under glancing angles of incidence. The doping effect is not limited to the immediate area of the surface tracks but affects SLG as a whole. After irradiation SLG is depleted of electrons and is effectively converted from an $n$-doped layer into a $p$-doped layer. The proposed mechanism, physical implantation of foreign atoms into the graphene lattice, can be used to achieve high doping levels without the global introduction of charged impurities which may act as scatterers. Our method could thus be very important for changing the carrier density in graphene devices with metal contacts, where the contact resistance limits the device performance and the ejection efficiency depends on the work function of SLG \cite{Song.2012}.

\section*{Acknowledgement}
We acknowledge financial support from the DFG in the frame work of the SPP 1459 {\it Graphene} (O.O.) and SFB 616: {\it Energy dissipation at surfaces} (B.K.B.) and from the European Community as an Integrating Activity Support of Public and Industrial Research Using Ion Beam Technology (SPIRIT) under EC contract no. 227012SPIRIT. The experiments were performed at the IRRSUD beamline of the Grand Accelerateur National d'Ions Lourds (GANIL), Caen, France. 

\section*{References}

\newpage

\end{document}